\documentclass[aps,prl,twocolumn,amsmath,amssymb,superscriptaddress,10pt,nobibnotes]{revtex4-2}
\pdfoutput=1

\usepackage{amsfonts, amssymb, amsmath, dsfont}
\usepackage{graphicx}
\usepackage{color}
\usepackage{ulem}
\usepackage[utf8]{inputenc}
\usepackage[T1]{fontenc}
\usepackage{mathrsfs}
\usepackage{hyperref}
\usepackage[svgnames]{xcolor}
\hypersetup{colorlinks,urlcolor=DarkBlue,linkcolor=DarkBlue,citecolor=DarkBlue,pdfdisplaydoctitle=true,pdfpagemode=UseOutlines,bookmarksnumbered=true,bookmarksopen=true} 
\usepackage{textcomp}

\renewcommand{\bm}[1]{\boldsymbol{\mathbf{#1}}}

\newcommand{\eq}[1]{Eq.~\eqref{#1}}

\newcommand{\fig}[1]{Fig.~\ref{#1}}

\newcounter{SIcount}

\newcommand{\beginsupplement}{%
        \setcounter{table}{0}
        \renewcommand{\thetable}{S\arabic{table}}%
        \setcounter{figure}{0}
        \renewcommand{\thefigure}{S\arabic{figure}}%
        \setcounter{section}{0}
        \renewcommand{\thesection}{\Roman{section}}
        \setcounter{equation}{0}
        \renewcommand{\theequation}{S\arabic{equation}}
     }

\begin{document}

\title{Optimizing illumination for precise multi-parameter estimations\\ in coherent diffractive imaging}

\author{Dorian Bouchet}
\altaffiliation{\href{mailto:dorian.bouchet@univ-grenoble-alpes.fr}{dorian.bouchet@univ-grenoble-alpes.fr}}
\affiliation{Nanophotonics, Debye Institute for Nanomaterials Science, Utrecht University, P.O. Box 80000, 3508 TA Utrecht, the Netherlands}
\affiliation{Present address: Université Grenoble Alpes, CNRS, LIPhy, 38000 Grenoble, France}
\author{Jacob Seifert}
\affiliation{Nanophotonics, Debye Institute for Nanomaterials Science, Utrecht University, P.O. Box 80000, 3508 TA Utrecht, the Netherlands}
\author{Allard P. Mosk}
\affiliation{Nanophotonics, Debye Institute for Nanomaterials Science, Utrecht University, P.O. Box 80000, 3508 TA Utrecht, the Netherlands}

\begin{abstract}
	Coherent diffractive imaging (CDI) is widely used to characterize structured samples from measurements of diffracting intensity patterns. We introduce a numerical framework to quantify the precision that can be achieved when estimating any given set of parameters characterizing the sample from measured data. The approach, based on the calculation of the Fisher information matrix, provides a clear benchmark to assess the performance of CDI methods. Moreover, by optimizing the Fisher information metric using deep learning optimization libraries, we demonstrate how to identify the optimal illumination scheme that minimizes the estimation error under specified experimental constrains. This work paves the way for an efficient characterization of structured samples at the sub-wavelength scale.
\end{abstract}

\maketitle

The fast and precise characterization of nanoscale devices is an essential aspect of advanced semiconductor manufacturing processes. It is thus crucial to ensure that optical measurements can reveal every important feature of nanostructured samples with an excellent precision. To achieve this goal, a common approach is to numerically reconstruct the permittivity distribution of the sample, either from interferometric measurements~\cite{haeberle_tomographic_2010} or from intensity measurements via ptychography-like techniques~\cite{rodenburg_ptychography_2019}. In many cases of interest, some \textit{a priori} knowledge of the sample is also available to the observer. For instance, in nanofabrication, the geometry of manufactured samples is usually known with high precision and only a few critical parameters need to be monitored after the lithography process~\cite{alexanderliddle_lithography_2011}. Typically, it is assumed that the sample can be described using a sparse representation in a known basis. Such an approach, referred to as sparsity-based CDI, leads to a significant reduction in the number of parameters that need to be estimated from the measured diffraction patterns, therefore mitigating ill-posedness of the inverse problem that needs to be solved~\cite{kamilov_optical_2016,liu_seagle_2018}. Furthermore, the resolution of reconstructed images is not limited by Rayleigh's criterion, so that parameters can be estimated with sub-wavelength precision~\cite{szameit_sparsity-based_2012,sidorenko_sparsity-based_2015,qin_deep_2016,zhang_far-field_2016}. 

As for any imaging technique, an important aspect of sparsity-based CDI is to identify an optimized approach to illuminate the sample~\cite{bian_content_2014,muthumbi_learned_2019}. Formally, the estimation precision achievable with different incident fields can be compared using the Cramér-Rao lower bound (CRLB), which is a central concept in estimation theory. This concept is currently widely used in single-molecule localization microscopy~\cite{deschout_precisely_2014,shechtman_optimal_2014} and in quantum metrology~\cite{szczykulska_multi-parameter_2016,sidhu_geometric_2020}. It has also been proposed as a new resolution measure for imaging systems~\cite{ram_beyond_2006,sentenac_influence_2007}, and the possibility to identify optimal incident fields that minimize the CRLB was recently investigated, for instance to localize a single particle in a complex environment~\cite{bouchet_influence_2020} or to characterize a phase object hidden behind a scattering medium~\cite{bouchet_maximum_2020}. 

\begin{figure}[ht]
	\centering
	\includegraphics[width=6.4cm]{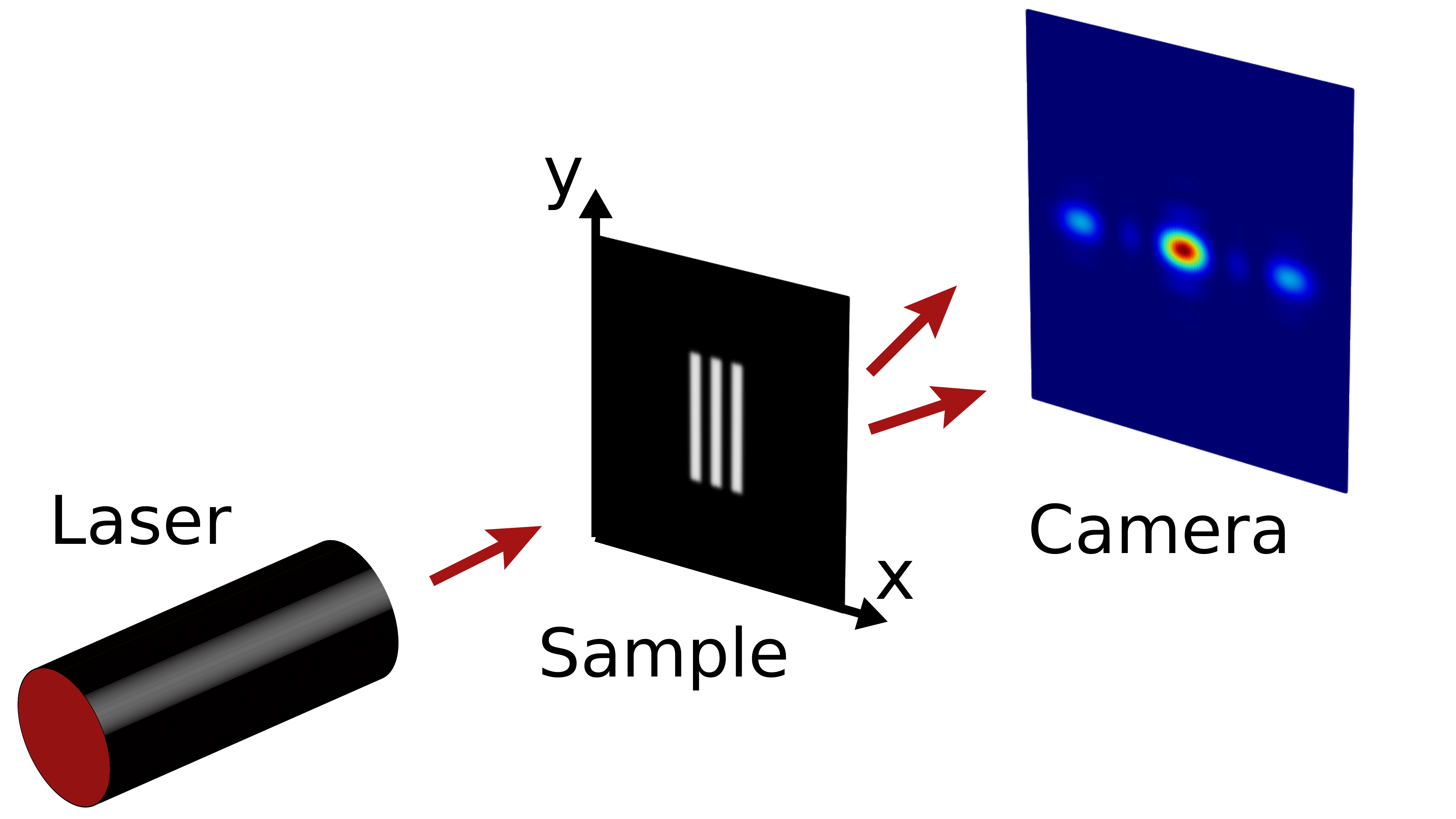}
	\caption{Representation of a CDI setup used for the characterization of a parameterized sample composed of three vertical lines. A coherent light source illuminates the sample, and diffraction patterns are measured by a camera located in the detection plane. }
	\label{fig1}
\end{figure}

In this Letter, we describe a method to find illumination schemes that optimize the precision of parameter estimation in sparsity-based CDI. As an example, we present different approaches to characterize a parameterized sample composed of three vertical lines (\fig{fig1}), either by determining optimal positions for the incident field or by identifying the optimal design for a zone plate that shapes the incident field. In addition, we analyze the resulting CRLB in terms of contributions of the quantum fluctuations of coherent states, the absence of phase information in the measurements and crosstalk between parameters. These results offer new insights to improve the performance of methods based on CDI when the dose per acquisition may be limited, notably for the characterization of delicate samples or when high throughput is required.

In CDI, one seeks to characterize a sample by estimating a set of $M$ parameters $\bm{\theta}=(\theta_1,\dots,\theta_M)$ from measurements of one or several diffraction patterns that constitute the data $\bm{X}$. Noise fluctuations in the data impose a fundamental limit to the achievable precision on the determination of $\bm{\theta}$. Indeed, the covariance matrix $\bm{\Sigma}$ of any unbiased estimator of $\bm{\theta}$ must satisfy the Cramér-Rao inequality, which states that the matrix $(\bm{\Sigma} - \bm{\mathcal{J}}^{-1})$ is always nonnegative definite~\cite{trees_detection_2013}. In this expression, the matrix $\bm{\mathcal{J}}$ is known as the Fisher information matrix, defined by $ \bm{\mathcal{J}} = \langle [ \bm{\nabla}_{\bm{\theta}} \ln p(\bm{X} ; \bm{\theta})] [ \bm{\nabla}_{\bm{\theta}} \ln p(\bm{X} ; \bm{\theta}) ]^\mathsf{T} \rangle$ where $p(\bm{X};\bm{\theta})$ is a joint probability density function, $\bm{\nabla}_{\bm{\theta}}$ is a partial derivative operator defined by $ \bm{\nabla}_{\bm{\theta}}= (\partial / \partial \theta_1, \dots,\partial / \partial \theta_{M})^\mathsf{T}$, and $\langle \dotsb \rangle $ denotes the expectation operator acting over noise fluctuations. While the probability density function $p(\bm{X};\bm{\theta})$ can describe any type of noise , we assume here that the values measured by the $N_\mathrm{p}$ pixels of the camera are statistically independent and follow a Poisson distribution, which corresponds to measurements limited only by shot noise. Considering a set of $N_\mathrm{m}$ diffraction patterns measured using different incident fields, the Fisher information matrix is then expressed by 
\begin{equation}
\left[\bm{\mathcal{J}}\right]_{ij}=\sum_{k,l} \frac{1}{I_{k,l}} \Biggl( \frac{\partial I_{k,l}}{\partial \theta_i} \Biggr)\left(\frac{\partial I_{k,l}}{\partial \theta_j} \right) \; ,
\label{information_matrix}
\end{equation}
where $I_{k,l}$ denotes the expected value of the intensity for the $k$-th pixel and for the $l$-th diffraction pattern. The resulting CRLB on the standard error on the estimated value of $\theta_i$ is given by
\begin{equation}
\mathcal{C}_i = \sqrt{\left[ \bm{\mathcal{J}}^{-1} \right]_{ii}} \; .
\label{crlb_full}
\end{equation}
This bound is asymptotically reached by maximum-likelihood (ML) estimators, which can be implemented by searching for the global maximum of the log-likelihood function~\cite{trees_detection_2013,thibault_maximum-likelihood_2012}. 

In conventional CDI, it is impractical to calculate the CRLB due to the computational complexity of inverting the large Fisher information matrices that arise when samples are described by many parameters~\cite{barrett_objective_1995,wei_cramer-rao_2020}. In contrast, the formalism is suitable to quantify the precision achievable with sparsity-based CDI, when samples can be described in sparse representations involving a reduced number of unknown parameters. In such cases, it is then possible to define an objective function that can be optimized to identify optimal illumination schemes tailored for the estimation of $\bm{\theta}$. For single-parameter estimations, the relevant objective function is simply given by the CRLB for the parameter~\cite{bouchet_maximum_2020}. For multi-parameter estimations, however, different relevant objective functions can be defined. As a possible objective function, one could choose the trace of $\bm{\mathcal{J}}^{-1}$, which provides a measure of the average CRLB but does not guarantee that a controlled threshold value bounds the CRLB for every parameter (see Supplementary Section 1). For this reason, we use the spectral radius of $\bm{\mathcal{J}}^{-1}$ as an objective function, which is defined as being the largest eigenvalue of $\bm{\mathcal{J}}^{-1}$. The CRLB on the standard error on the estimated value of the first principal component is then expressed as follows:
\begin{equation}
\mathcal{C}_\rho = \sqrt{ \rho \left(\bm{\mathcal{J}}^{-1}\right) } \; ,
\end{equation}
where $\rho(\bm{\mathcal{J}}^{-1})$ denotes the spectral radius of $\bm{\mathcal{J}}^{-1}$. The inequality $\mathcal{C}_i \leq \mathcal{C}_\rho$ holds for any parameter $\theta_i$. Minimizing this objective function essentially leads to a reduction of the CRLB for the parameters that are the most difficult to estimate, a feature that is highly desirable for practical applications when the metrological specifications involve a single tolerance value that applies to all parameters.

To demonstrate the benefits of this approach in sparsity-based CDI, we consider a sample composed of three vertical lines (\fig{fig1}). These lines are separated from each other by a distance of $10$\,\textmu m, each line being characterized by a width of $10$\,\textmu m and a length of $100$\,\textmu m. A sparse representation of the sample is obtained by describing these lines with $12$ parameters $\bm{\theta}=(x_1,\dots,x_6,y_1,\dots,y_6)$, corresponding to the coordinates of the edges of the lines. We assume that the sample is illuminated with a coherent field at a wavelength $\lambda=561$\,nm. We choose a total number of photons incident on the sample of $n=3 \times 10^6$; one can deduce the CRLB for other values of $n$ by remarking that the CRLB for shot-noise limited measurements scales with $1/\sqrt{n}$. Diffraction patterns are calculated using a scalar diffraction approach by propagating the resulting field using the angular spectrum representation. This method allows us to calculate the expected value of the intensity $I_{k,l}$ that would be measured by a camera located at a distance $z=10$\,mm from the sample, and thus to calculate the associated $12\times12$ Fisher information matrix using a finite-difference approximation of \eq{information_matrix} (see Supplementary Section 2).

\begin{figure}[t]
	\centering
	\includegraphics[width=\linewidth]{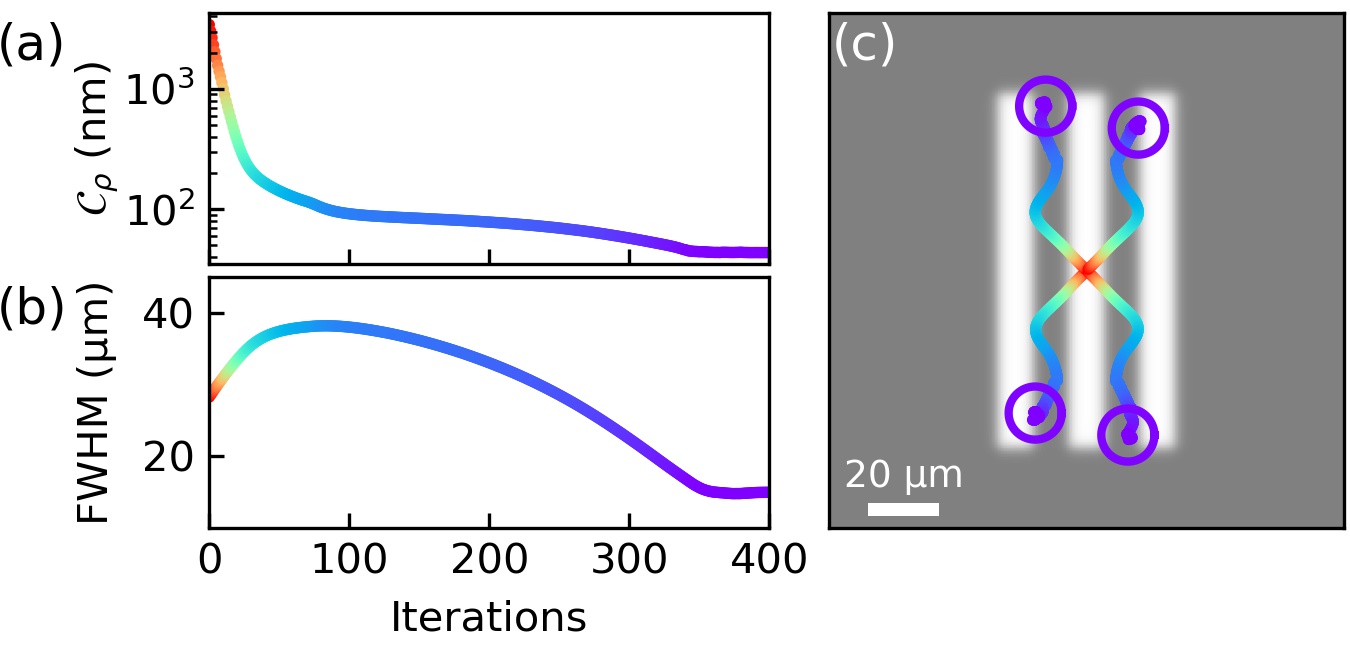}
	\caption{Evolution of (a)~the objective function $\mathcal{C}_\rho$ during the optimization process, as well as (b)~the FWHM of the Gaussian probe field and (c)~the probe positions represented in the sample plane. The purple circles shown in~(c) are centered at the optimized positions, with a diameter equal to the optimized FWHM. The color of the curves represents the value of $\mathcal{C}_\rho$, ranging from $3.4$\,\textmu m (red) to $44$\,nm (purple).}
	\label{fig2}
\end{figure}

\begin{figure}[t]
	\centering
	\includegraphics[width=\linewidth]{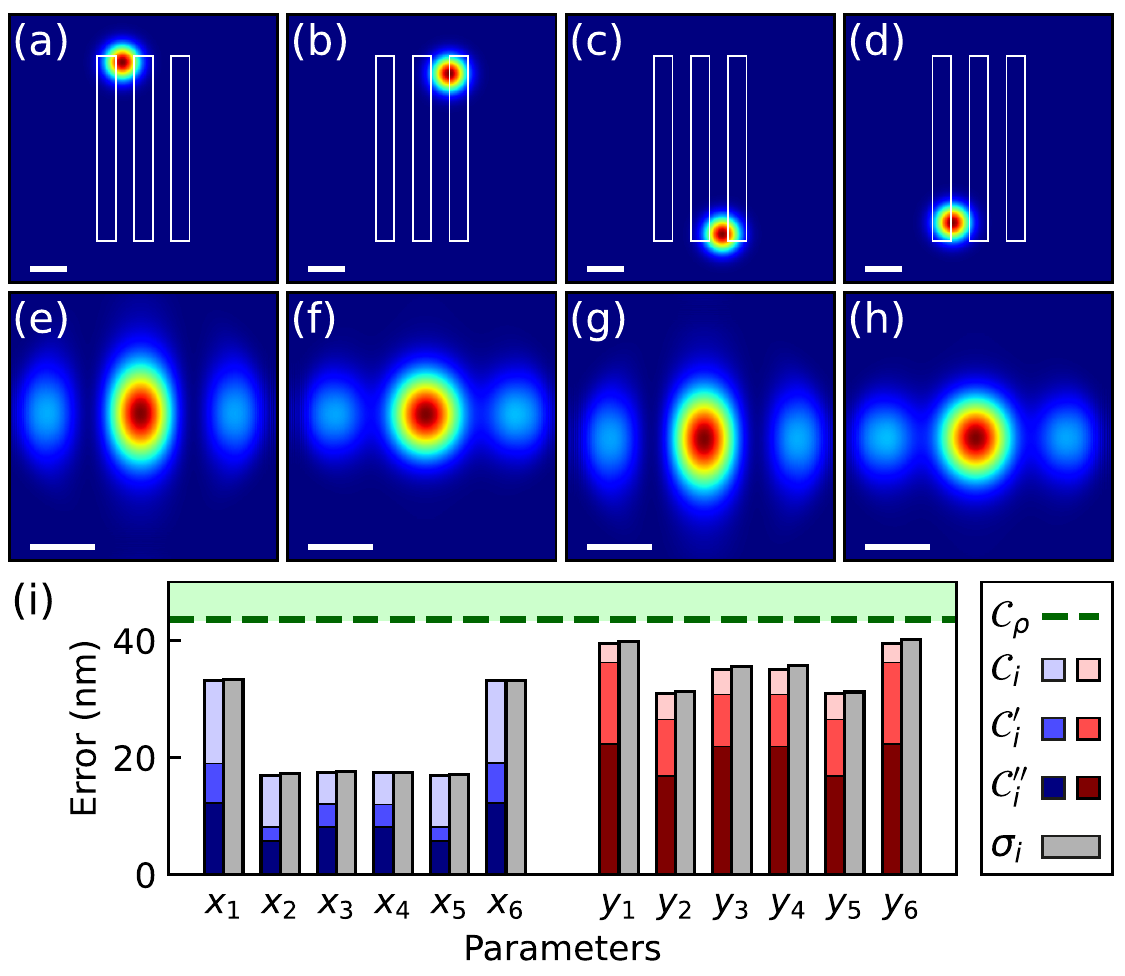}
	\caption{(a--d)~Spatial distributions of the excitation intensity in the sample plane for the optimal probe positions, assuming that four diffraction patterns are measured. The position of the sample is represented by white lines. Scale bars, $20$\,\textmu m. (e--h) Spatial distributions of the intensity in the detection plane. Scale bars, $200$\,\textmu m. (i)~CRLB for each parameter after the minimization of $\mathcal{C}_\rho$, along with the RMS error obtained by performing ML estimations on $10^4$ numerically-generated diffraction patterns.}
	\label{fig3}
\end{figure}

Tailoring the spatial distribution of the probe field provides us with degrees of freedom that can be tuned to minimize $\mathcal{C}_\rho$. In a constrained configuration, the shape of the distribution is fixed (e.g. a Gaussian beam) and it is desired to identify optimal values for the position of the probe field and its spatial extent. To solve this optimization problem, we employ the Adam optimizer, which is commonly used to train deep neural networks~\cite{kingma_adam_2015,barbastathis_use_2019} and which is implemented in the open-source platform TensorFlow. 
We first consider the acquisition of four independent diffraction patterns, each of them obtained by illuminating the sample using a Gaussian beam with $n/4$ photons. The Adam optimizer is then used to identify the probe positions and the full width at half maximum (FWHM) that minimize the CRLB for the first principal component $\mathcal{C}_\rho$. Note that such optimization procedure is especially effective when the \textit{a priori} knowledge available on $\bm{\theta}$ is of the order of the FWHM of the probe field (see Supplementary Section 3). After the optimization process, the value of $\mathcal{C}_\rho$ is $44$\,nm (\fig{fig2}a), which is well below the wavelength of the incident light thanks to the sparse representation of the object. Optimal probe positions are identified at critical areas of the sample, with an optimized FWHM of $15$\,\textmu m (\fig{fig2}b,c, \fig{fig3}a--h). This optimal illumination scheme can be interpreted as a trade-off between the necessity to illuminate all important areas of the object and the requirement to minimize the number of photons wasted by missing the object or the camera. For comparison, we performed the same analysis for a conventional ptychographic scheme. To ensure that the probes significantly overlap over the field of view~\cite{bunk_influence_2008}, we chose a FWHM of $100$\,\textmu m and four probe positions distributed in a square grid of side length $50$\,\textmu m centered on the object. The value of $\mathcal{C}_\rho$ obtained with this conventional scheme is $127$\,nm, hence showing that $\mathcal{C}_\rho$ is reduced by a factor of $3$ with the optimized scheme. 

We can also use \eq{crlb_full} calculate the CRLB for each parameter after the minimization of $\mathcal{C}_\rho$ (\fig{fig3}i). Interestingly, the formalism allows us to analyze the contribution of different error sources. Indeed, information is partly lost both because of the influence of parameter crosstalk and because the phase of the field $\varphi_{k,l}$ is not captured by the measurements. When $\theta_i$ is to be estimated, other parameters can be considered as nuisance parameters that can increase the CRLB via crosstalk~\cite{trees_detection_2013}. Estimations of $\theta_i$ are the same regardless of whether other parameters are known or unknown only if $[\bm{\mathcal{J}}]_{ij}=0$ for $i \neq j$. We can thus assess the influence of parameter crosstalk by calculating the lower bound on the standard error on the estimated value of $\theta_i$ as if the Fisher information matrix was diagonal. This bound is given by $\mathcal{C}'_i=1/\sqrt{\mathcal{J}'_i}$, where 
\begin{equation}
\mathcal{J}'_i = \sum_{k,l} \frac{1}{I_{k,l}} \left( \frac{\partial I_{k,l}}{\partial \theta_i} \right)^2 \; .
\end{equation}
In addition, the absence of phase measurements also leads to an increase of the CRLB. This can be assessed by calculating the lower bound on the standard error on the estimated value of $\theta_i$ assuming that both the intensity and the phase can be measured by the observer -- the precision of estimations is then only limited by the quantum fluctuations of coherent states. This bound is expressed by $\mathcal{C}''_i=1/\sqrt{\mathcal{J}''_i}$, where $\mathcal{J}''_i = 4 \sum |\partial E_{k,l}/\partial \theta_i|^2$ is the Fisher information corresponding to single-parameter estimation using an ideal homodyne detection scheme~\cite{bouchet_maximum_2020}. Note that $\mathcal{J}''_i$ is also equal to the quantum Fisher information associated with the estimation of a single parameter from uncorrelated coherent states~\cite{helstrom_quantum_1969,bouchet_maximum_2020}. Introducing $E_{k,l}=\sqrt{I_{k,l}} \exp(i \varphi_{k,l})$, we can decompose $\mathcal{J}''_i$ as follows:
\begin{equation}
\mathcal{J}''_i = \sum_{k,l} \frac{1}{I_{k,l}} \left( \frac{\partial I_{k,l}}{\partial \theta_i} \right)^2 + 4 \sum_{k,l} I_{k,l} \left( \frac{\partial \varphi_{k,l}}{\partial \theta_i} \right)^2\; .
\label{fisher_amplitude_phase}
\end{equation}
The two terms that appear in the second member of \eq{fisher_amplitude_phase} can be interpreted as the Fisher information enclosed in the intensity and the phase of the detected field, respectively.

The different bounds that are introduced here satisfy the chain of inequalities $\mathcal{C}''_i \leq \mathcal{C}'_i \leq \mathcal{C}_i \leq \mathcal{C}_\rho $, as can be seen in \fig{fig3}i. The influence of parameter crosstalk varies depending on the considered parameter, but we observe that parameters defining the $x$-position of the line edges are more affected than those defining the $y$-position of the line edges. Furthermore, after the propagation of the field to the detection plane, the Fisher information associated with intensity and phase measurements (first and second terms of the second member of \eq{fisher_amplitude_phase}, respectively) are approximately equal, which explains why the CRLB is then degraded by a factor close to $\sqrt{2}$ by the absence of phase information. 

In order to show that the calculated CRLB can be approached with ML estimators, we numerically generate a set of $10^4$ noisy diffraction patterns. For each pattern, we first randomly modify the value of each parameter according to a normal distribution, with a standard deviation of $0.5$\,\textmu m. We then calculate the expected value of the intensity in the detection plane, and use it to randomly generate noisy data with Poisson statistics. The value of all parameters is then estimated by maximizing the log-likelihood function with the Adam optimizer. The root-mean square (RMS) error $\sigma_i$ of the estimated values of each parameter is close to the fundamental limit $\mathcal{C}_i$ (\fig{fig3}i), which demonstrates here the efficiency of the ML estimator.

\begin{figure}[t]
	\centering
	\includegraphics[width=\linewidth]{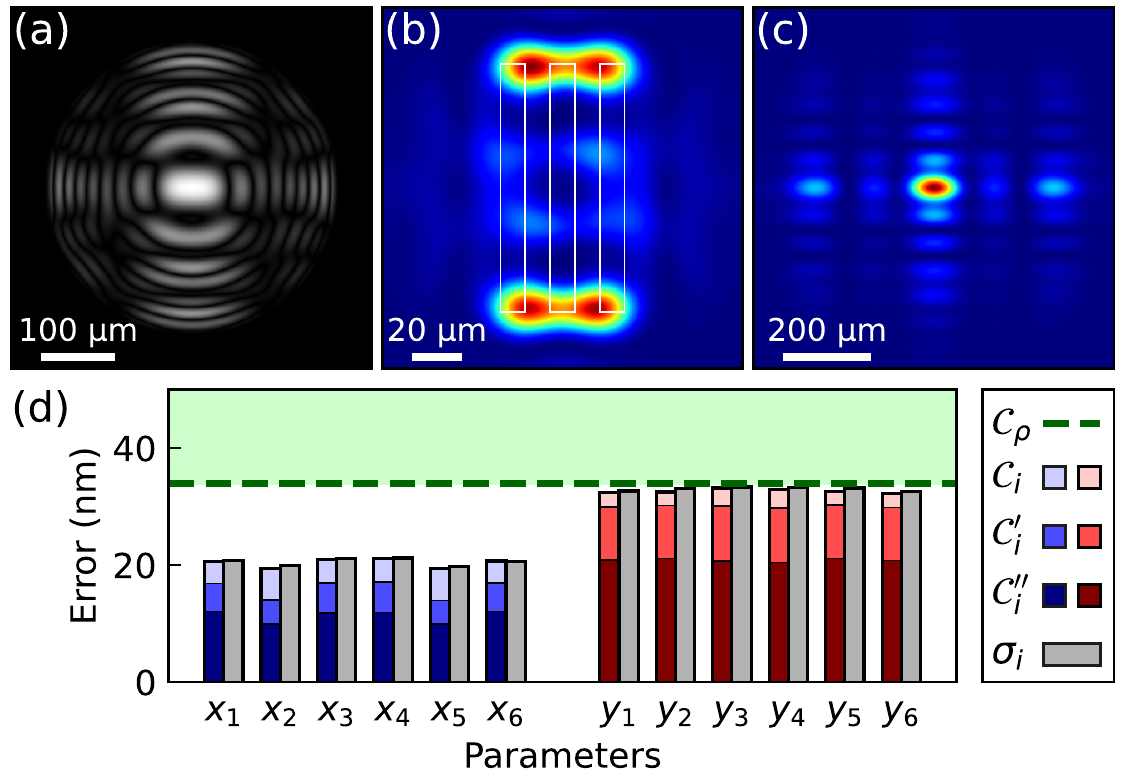}
	\caption{(a)~Optimal design of the zone plate that minimize $\mathcal{C}_\rho$, assuming that a single diffraction pattern is measured. (b)~Spatial distribution of the excitation intensity in the sample plane, as generated by the optimized zone plate. The position of the sample is represented by white lines. (c)~Resulting spatial distribution of the intensity in the detection plane. (d)~CRLB for each parameter after the minimization of $\mathcal{C}_\rho$, along with the RMS error obtained by performing ML estimations on $10^4$ numerically-generated diffraction patterns.}
	\label{fig4}
\end{figure}

It is known that a structured illumination can improve the resolution of imaging techniques, which notably led to the development of randomized zone plates for use in ptychography~\cite{morrison_x-ray_2018,odstrcil_towards_2019}. Here, we can use our numerical framework to deterministically identify the design of the zone plate that is optimal for precisely characterizing the sample. To this end, we now consider a continuous transmission mask located at a distance of $10$\,mm upstream of the sample. The radius of the zone plate is set to $180$\,\textmu m, so that the largest spatial frequency of the field in the sample plane is the same as for the Gaussian beams represented in \fig{fig3}a--d. Starting from a uniform initial guess, we run the Adam optimizer to find the design of the zone plate that minimizes $\mathcal{C}_\rho$ for a single-shot measurement (\fig{fig4}a). This zone plate generates an intensity in the sample plane that is high at all critical areas of the sample (\fig{fig4}b), producing a structured intensity pattern in the detection plane (\fig{fig4}c). As shown in \fig{fig4}d, the value of $\mathcal{C}_\rho$ resulting from the optimization process is $34$\,nm, which is significantly lower than the optimized value of $44$\,nm obtained in the case of the Gaussian beams. Thus, for a given total number of photons incident on the sample, a single-shot measurement using the optimized zone plate allows for a better precision on the estimation of $\bm{\theta}$ as compared to what can be achieved with four measurements performed using a Gaussian beam illuminating the sample at different positions. This demonstrates the potential of optimized zone plates for the precise characterization of structured sampled at high throughput, as often needed for industrial applications~\cite{alexanderliddle_lithography_2011}.

In summary, we calculated the CRLB to assess the precision achievable with sparsity-based CDI, and we used the formalism to identify optimal illumination schemes that allow all parameters to be precisely estimated while limiting the number of photons interacting with the sample.
We envision that this strategy could be applied in future work by representing objects with different choices of basis functions, such as a wavelet basis or a basis of Gabor functions~\cite{barrett_foundations_2013}. Implementing a Bayesian approach could also allow for more flexibility in the \textit{a priori} knowledge that can be described using the formalism~\cite{trees_detection_2013}. Furthermore, advanced numerical frameworks could be used to go beyond the first Born approximation and to characterize strongly scattering samples in two or three dimensions~\cite{rodenburg_ptychography_2019,dilz_domain_2017}.

\bigskip

\noindent{\textbf{Funding.}} 
Netherlands Organization for Scientific Research NWO (Vici 68047618 and Perspective P16-08).

\bigskip

\noindent{\textbf{Acknowledgments.}} The authors thank W. Coene and L. Loetgering for insightful discussions and C. de Kok for IT support.

\bigskip

\noindent{\textbf{Disclosures.}} The authors declare no conflicts of interest.



\onecolumngrid
\clearpage
\beginsupplement
\begin{center}
	\textbf{\large Optimizing illumination for precise multi-parameter estimations\\ in coherent diffractive imaging\\ \vspace{0.5cm} Supplementary Information}
	
	\bigskip
	Dorian Bouchet,$^{1,2}$ Jacob Seifert,$^1$ and Allard P. Mosk$^1$\\ \vspace{0.15cm}
	\textit{\small $^\mathit{1}$Nanophotonics, Debye Institute for Nanomaterials Science,\\ Utrecht University, P.O. Box 80000, 3508 TA Utrecht, the Netherlands}\\
	\textit{\small $^\mathit{2}$Present address: Université Grenoble Alpes, CNRS, LIPhy, 38000 Grenoble, France}\\	
	
\end{center}
\vspace{1cm}
\twocolumngrid

\section{I. Choice of the objective function}

Finding optimal incident fields requires one to clearly define a criterion for optimality, which needs to be attached to a scalar quantity for this criterion to constitute a suitable objective function.  In the single-parameter case, the Fisher information is already a scalar quantity and minimizing the CRLB (calculated as the inverse of the Fisher information) constitutes a straightforward objective function~\cite{bouchet_maximum_2020_2}. In contrast, for multiple parameter estimations, the Fisher information is a matrix, and the scalar quantity that is to be constructed from this matrix depends on the metrological specifications imposed on the precision that must be achieved in the estimation of each parameter.

Whenever the variance of the estimates averaged over all parameter needs to be minimized, a relevant objective function is constituted by the trace of the inverse of the Fisher information matrix, noted $\operatorname{Tr}(\bm{\mathcal{J}}^{-1})$. This criterion is invariant under orthonormal transformations, which is a desirable property since it ensures that the optimization procedure yields the same optimal incident field for two equivalent parameterizations. However, this objective function does not guarantee that a controlled threshold value bounds the CRLB for every parameter. Instead, a high CRLB for one parameter can in principle be compensated by a low CRLB for other parameters, since the objective function is constructed as a sum of the $\mathcal{C}_i^2$. This is not a desirable feature for several practical applications, when a tolerance is specified and when all parameters need to be estimated with a precision that is equal or better than the tolerance.  

For this reason, in the manuscript, we opted for a different objective function constituted by the spectral radius of the inverse of the Fisher information matrix, noted $\rho(\bm{\mathcal{J}}^{-1})$. This criterion is also invariant under orthonormal transformations (as opposed for instance to an objective function that would be defined as the largest diagonal coefficient of $\bm{\mathcal{J}}^{-1}$). Moreover, it directly yields an upper bound that applies to the CRLB of all parameters. Thus, it allows one to easily specify a single tolerance associated with the maximum error that can be made in estimating every parameters. 

\begin{figure}[ht]
	\centering
	\includegraphics[width=\linewidth]{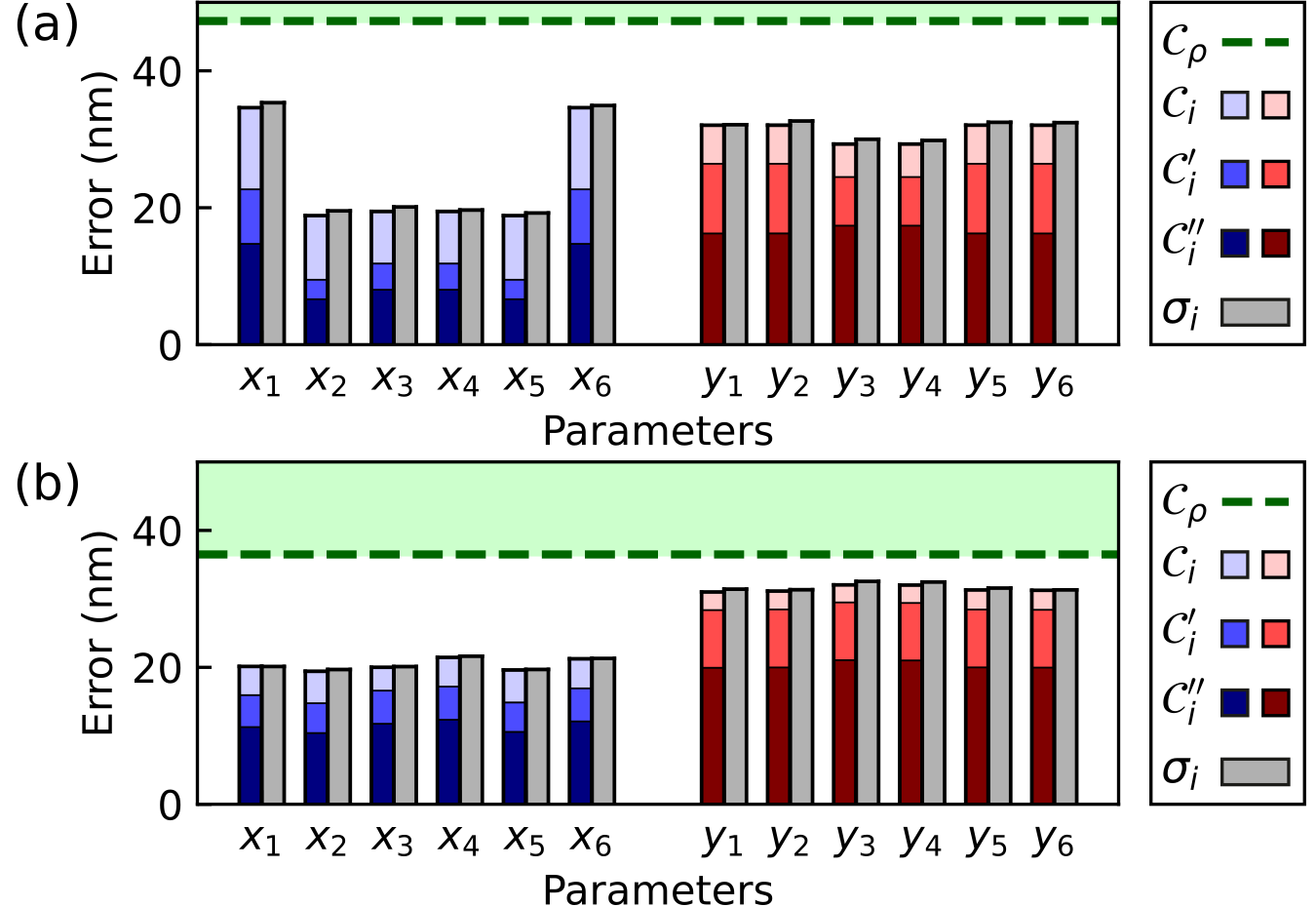}
	\caption{CRLB for each parameter after the minimization of $\operatorname{Tr}(\bm{\mathcal{J}}^{-1})$ along with the RMS error obtained by performing ML estimations on $10^4$ numerically-generated diffraction patterns, in the case of a)~the optimization of the positions of four Gaussian beams and b)~the optimization of a zone plate located upstream of the sample. These figures can be compared to Fig.~3i and Fig.~4d of the manuscript, which show the same results after a minimization of $\rho(\bm{\mathcal{J}}^{-1})$.
	}
	\label{figS1}
\end{figure}

In order to compare the results obtained by minimizing $\rho(\bm{\mathcal{J}}^{-1})$ and $\operatorname{Tr}(\bm{\mathcal{J}}^{-1})$, we calculated the CRLB for each parameter after the minimization of $\operatorname{Tr}(\bm{\mathcal{J}}^{-1})$ in the case of an optimization of the positions of four Gaussian beams (Fig.~\ref{figS1}a) and in the case of an optimization of the zone plate located upstream of the sample (Fig.~\ref{figS1}b). It clearly appears that, in both cases, the resulting CRLB for each parameter is very close to the CRLB obtained by minimizing $\rho(\bm{\mathcal{J}}^{-1})$ (see Fig.~3i and Fig.~4d of the manuscript). Nevertheless, the CRLB for the first principal component $\mathcal{C}_\rho$ is slightly higher when $\operatorname{Tr}(\bm{\mathcal{J}}^{-1})$ is minimized. Indeed, the value of $\mathcal{C}_\rho$ is $47$\,nm (instead of $44$\,nm) when the four Gaussian beams are optimized, and the value of $\mathcal{C}_\rho$ is $36$\,nm (instead of $34$\,nm) when the zone plate is optimized. Thus, assuming that the metrological specifications involve a single tolerance value that applies to all parameters, there is here a slight disadvantage of minimizing $\operatorname{Tr}(\bm{\mathcal{J}}^{-1})$ instead of $\rho(\bm{\mathcal{J}}^{-1})$.

\section{II. Numerical methods}

We implemented a numerical model based on scalar wave propagation at a wavelength of $\lambda=561\,$nm. In the detection plane ($z=z_{\mathrm{det}}$), the field $E^{\mathrm{det}}_l(x',y')$ associated with the $l$-th measured diffraction pattern is represented by a $128 \times 128$ complex-valued array, with a pixel size of $6.45\,$\textmu m. The field in the object plane ($z=0$) is oversampled by a factor of $8$ and thereby represented by a $1024\times 1024$ complex-valued array. The distance between the object and the camera is assumed to be $z_{\mathrm{det}}=10\,$mm. The object under consideration $O(x,y)$ is composed of three vertical lines, described with $12$ parameters $\bm{\theta}=(x_1,\dots,x_6,y_1,\dots,y_6)$ that correspond to the coordinates of the edges of the lines. The numerical approach that we employ to evaluate the Fisher information matrix requires that the $1024\times 1024$ array representing the object function is differentiable with respect to $\bm{\theta}$. For this reason we constructed each line by multiplying four sigmoid functions ranging between $0$ and $1$. With this strategy, the components of $\bm{\theta}$ are then defined as being the coordinates for which the sigmoid functions take the value $1/2$.

This object function $O(x,y)$ is multiplied by the incident field $E^{\mathrm{inc}}_l(x,y)$ evaluated in the object plane. Within the projection approximation, this procedure yields a correct estimate of the transmitted field $E^{\mathrm{obj}}_l(x,y)$ in the object plane. Introducing the wavenumber $k_0=2 \pi/\lambda$ and using the angular spectrum representation of plane waves~\cite{goodman_introduction_2017_2}, we can calculate the field in the detection plane as follows:
\begin{equation}
\begin{split}
E^{\mathrm{det}}_l(x',y')  = \frac{1 }{4 \pi^2} \iint  \tilde{E}^{\mathrm{obj}}_l(\alpha,\beta) \exp \left(i \gamma z_{\mathrm{det}}\right) \\
 \exp\left[i (\alpha x'+ \beta y')\right] \mathrm{d} \alpha \mathrm{d} \beta \; ,
\label{field_camera}
\end{split}
\end{equation}
where we noted $\gamma=\sqrt{k_0^2-\alpha^2-\beta^2}$ and where $\tilde{E}^{\mathrm{obj}}_l(\alpha,\beta)$ is the Fourier transform of $E^{\mathrm{obj}}_l(x,y)$ expressed by
\begin{equation}
\begin{split}
\tilde{E}^{\mathrm{obj}}_l(\alpha,\beta) = \iint E^{\mathrm{inc}}_l(x,y) O(x,y) \\
 \exp \left[-i(\alpha x + \beta y) \right] \mathrm{d} x \mathrm{d} y \; .
\label{fourier_field_object}
\end{split}
\end{equation}
The Fourier transform operations involved in Eqs.~(\ref{field_camera}) and~(\ref{fourier_field_object}) can be numerically implemented with a fast Fourier transform (FFT) algorithm. Nevertheless, in order to avoid ringing artifacts, $\tilde{E}^{\mathrm{obj}}_l(\alpha,\beta)$ is first multiplied by a circular aperture function before integration in Eq.~(\ref{field_camera}). The radius of this aperture function is determined from the effective numerical aperture of the detection apparatus, and we convolve this aperture function with a 2-dimensional Hann function in order to avoid a hard truncation of the field in the frequency domain. Once the field $E^{\mathrm{det}}_l(x',y')$ in the detection plane is calculated according to Eq.~(\ref{field_camera}), the intensity in this plane is then simply expressed by $I_{k,l}=|E^{\mathrm{det}}_l(x_k',y_k')|^2$. The Fisher information matrix is then numerically estimated using $\bm{\mathcal{J}} \simeq \bm{H}^\mathsf{T} \bm{H}$, where
\begin{equation}
[\bm{H}]_{ki} = \sum_{l}  \left[ \frac{I_{k,l}(\theta_i + \Delta \theta)-I_{k,l}(\theta_i - \Delta \theta)}{2 \Delta \theta \sqrt{I_{k,l}(\theta_i)+\epsilon_\mathrm{r}}} \right] \; .
\end{equation}
All results presented in this work are obtained using a step size $\Delta \theta=1$\,nm and a regularization parameter $\epsilon_\mathrm{r}=0.01$, which has the physical interpretation of being the expected value of an additive noise with Poisson statistics. Results are then found to be insensitive to $\Delta \theta$ and $\epsilon$ over several orders of magnitude, attesting that these values yield here an accurate estimate of the Fisher information matrix.

The optimization was performed using a NVIDIA GeForce RTX 2070, which is a commercial graphics processing unit (GPU). The optimization procedure relies on the Adam optimizer~\cite{kingma_adam_2015_2}, called from TensorFlow libraries. This optimizer takes one main hyperparameter (the learning rate) which must be determined heuristically. We found that learning rates between $0.3$ and $0.4$ are appropriate to efficiently minimize the objective function, defined as the logarithm of the largest eigenvalue of the inverse of the Fisher information matrix. We used the default values provided by TensorFlow for other the hyperparameters of the Adam optimizer ($\beta_1=0.9$, $\beta_2=0.999$ and $\epsilon=10^{-7}$). 

At first, we have found the optimized fields that minimize $\mathcal{C}_\rho$ by tuning the positions of four Gaussian beams and their spatial extent. The full width at half maximum (FWHM) was initialized at $28$\,\textmu m, and the coordinates of the probes were initialized at $\pm$0.4\,\textmu m (see Fig.~2 of the manuscript). Using the Adam optimizer, we performed $400$ iterations with a learning rate of $0.4$ to simultaneously optimize the probe positions and the FWHM of the probes. On our GPU, this was achieved in a time of $149$\,s.

Then, we have found the optimized zone plate that minimize $\mathcal{C}_\rho$ by tuning the amplitudes of the $1024\times1024$ array representing a zone plate located upstream of the sample. We considered a zone plate with a radius of $180$\,\textmu m, and we supposed that the distance between the zone plate and the sample is $10$\,mm. The amplitudes defining the design of the zone plate were initialized with random values taken from a uniform distribution. Using the Adam optimizer, we performed $300$ iterations with a learning rate of $0.3$ to optimize these amplitudes. On our GPU, this was achieved in a time of $32$\,s. Note that this time is significantly lower than the time needed to find the optimized probe positions. This difference arises from the fact that four Fisher information matrices need to be evaluated per iteration to find optimal probe positions (one for each probe position), whereas only one Fisher information matrix needs to be evaluated per iteration to identify the optimized zone plate.

\section{III. Influence of an inaccurate prior knowledge}

In general, the Fisher information matrix depends on the value taken by all parameters $\bm{\theta}$ that describe the object. Consequently, optimal illumination schemes depend on the values of $\bm{\theta}$ that were assumed during the optimization process. These values must be inferred from an \textit{a priori} knowledge of the object, which can be obtained for instance through design considerations or using a low-intensity plane-wave illumination. In order to test what kind of \textit{a priori} knowledge of the object is required for the optimization process to be effective, we study here the precision that can be achieved on the estimation of the parameter $y_2$ (top edge of the left line) for different illumination schemes. In the following, the origin of the coordinate system in the object plane $(x=0,y=0)$ is defined as being located in the center of the middle line. 

\begin{figure}[ht]
	\centering
	\includegraphics[width=\linewidth]{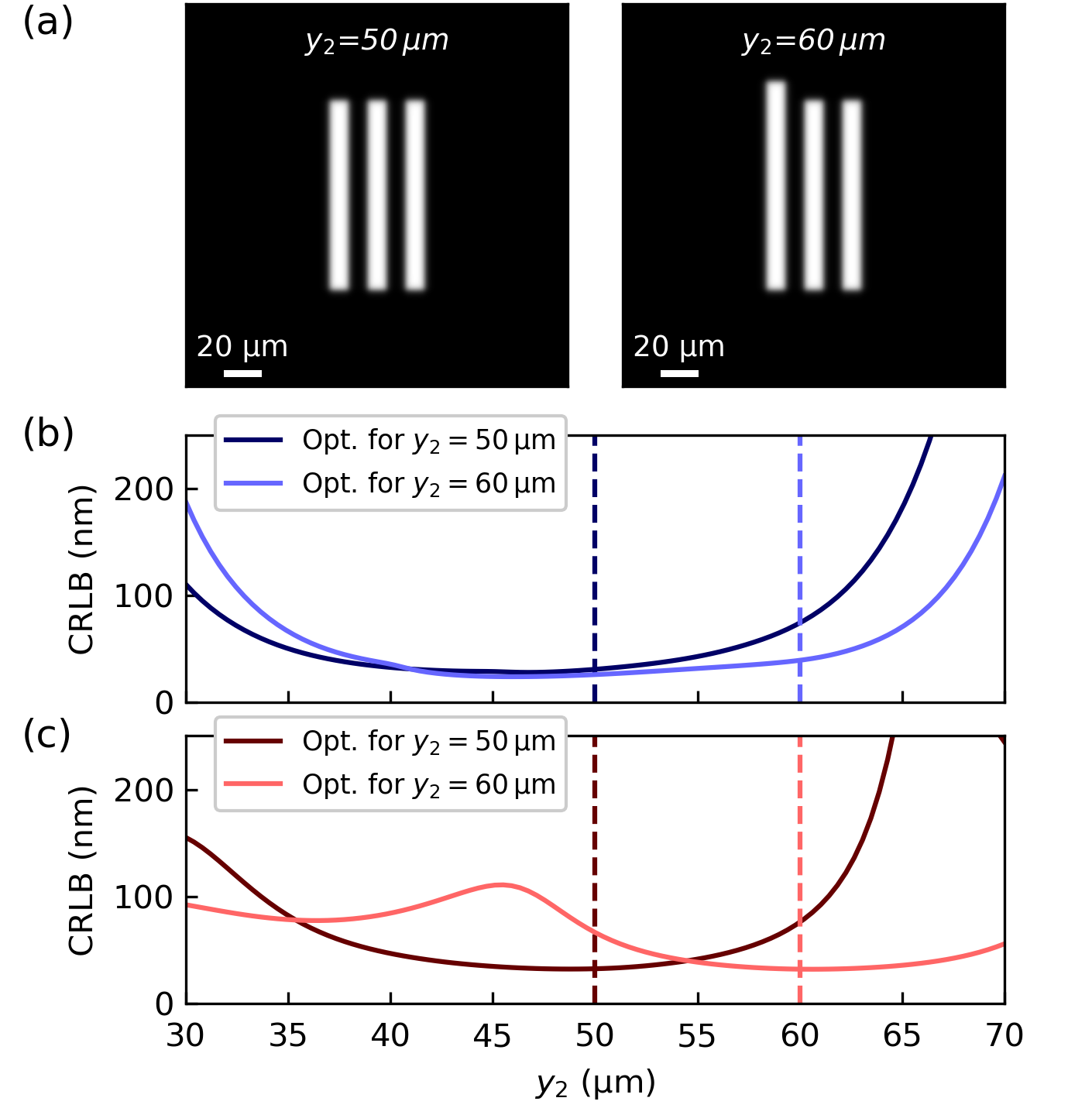}
	\caption{a)~Object function under the hypothesis that $y_2=50$\,\textmu m (left) and under the hypothesis that $y_2=60$\,\textmu m (right). b)~CRLB for the parameter $y_2$ as a function of the true value taken by this parameter in the case of the optimization of the positions of four Gaussian beams. c)~CRLB for the parameter $y_2$ in the case of the optimization of a zone plane located upstream the sample. Opt. stands for Optimized (dark and light curves represent the CRLB obtained after an optimization procedure performed under the hypothesis $y_2=50\,$\textmu m and $y_2=60\,$\textmu m, respectively).}
	\label{figS2}
\end{figure}

We first consider the illumination scheme involving four Gaussian probes that minimizes $\mathcal{C}_\rho$ under the hypothesis that $y_2=50$\,\textmu m, which is equivalent to a total line length of $100$\,\textmu m and corresponds to the situation considered in the manuscript (Fig.~\ref{figS2}a, left). For this illumination scheme, we vary the true value taken by $y_2$ and we calculate the CRLB associated with the estimation of this parameter (Fig.~\ref{figS2}b, dark blue curve). We observe that the CRLB is minimized when the true value of $y_2$ matches the value assumed during the optimization process ($y_2=50$\,\textmu m), and that the CRLB remains close to this minimum value within a range of the order of the FWHM of the probe field ($15$\,\textmu m). For larger variations of $y_2$, the critical area of the sample that depends on this parameter is not properly illuminated by the incident field, resulting in a higher CRLB. For comparison purposes, we also identify the illumination scheme involving four Gaussian probes that minimizes $\mathcal{C}_\rho$ under the hypothesis that $y_2=60$\,\textmu m (Fig.~\ref{figS2}a, right). For this new illumination scheme, the CRLB is minimized for this value of $y_2$ and remains close to this minimum value within a relatively large range (Fig.~\ref{figS2}b, light blue curve), which confirms that the optimization procedure is here robust with respect to an imperfect \textit{a priori} knowledge of the sample.

For completeness, we perform the same analysis for the illumination scheme that involves a zone plate minimizing $\mathcal{C}_\rho$, as described in the manuscript (see Fig.~4 of the manuscript). Similarly, we observe that the CRLB is minimized when the true value of $y_2$ matches the value assumed during the optimization process, which is either $y_2=50$\,\textmu m (Fig.~\ref{figS2}c, dark red curve) or $y_2=60$\,\textmu m (Fig.~\ref{figS2}c, light red curve). Note that all optimized fields that we identified here are optimally shaped for the simultaneous estimation of all $12$ parameters describing the object. The complex shape of the resulting illumination patterns in the object plane (see for instance the intensity distribution shown in Fig.~4b of the manuscript) can then give rise to a non-convex dependence of the CRLB upon the value taken by the parameters, as observed in Fig.~\ref{figS2}c (light red curve).

\end{document}